\documentclass[compsoc,conference,a4paper,10pt,times]{IEEEtran}
\usepackage[compatibility=false]{caption}
\IEEEoverridecommandlockouts

\usepackage{amsmath,amssymb,amsfonts}
\usepackage{algorithmicx}
\usepackage{algpseudocode}
\usepackage{graphicx}
\usepackage{textcomp}
\usepackage{bmpsize}
\usepackage{enumitem}
\usepackage{xcolor}
\usepackage{lipsum}
\usepackage{listings}
\usepackage{booktabs}
\usepackage{subcaption}
\usepackage{algorithm}
\usepackage{amssymb}
\usepackage{algpseudocode}
\usepackage{amsmath,amssymb,stmaryrd}
\usepackage{xspace}
\usepackage{pgfplots}
\pgfplotsset{compat=1.18}
\usepgfplotslibrary{groupplots}

\usepackage[colorlinks=true,urlcolor=black,linkcolor=blue]{hyperref}

\def\BibTeX{{\rm B\kern-.05em{\sc i\kern-.025em b}\kern-.08em
    T\kern-.1667em\lower.7ex\hbox{E}\kern-.125emX}}
\begin{document}

\title{Analyzing Chain of Thought (CoT) Approaches in Control Flow Code Deobfuscation Tasks }

\author{
\IEEEauthorblockN{Seyedreza Mohseni\IEEEauthorrefmark{1},
Sarvesh Baskar\IEEEauthorrefmark{1},
Edward Raff\IEEEauthorrefmark{2}\IEEEauthorrefmark{1},
Manas Gaur\IEEEauthorrefmark{1}}
\IEEEauthorblockA{\IEEEauthorrefmark{1}\textit{University of Maryland Baltimore County}, Baltimore, USA}
\IEEEauthorblockA{mohseni1@umbc.edu, sarvesh@umbc.edu, manas@umbc.edu}
\IEEEauthorblockA{\IEEEauthorrefmark{2}\textit{CrowdStrike}, New York, USA, edward.raff@crowdstrike.com}
}

\maketitle

\begin{abstract}

Code deobfuscation is the task of recovering a readable version of a program while preserving its original behavior. In practice, this often requires days or even months of manual work with complex and expensive analysis tools. In this paper, we explore an alternative approach based on Chain-of-Thought (CoT) prompting, where a large language model is guided through explicit, step-by-step reasoning tailored for code analysis. We focus on control flow obfuscation, including Control Flow Flattening (CFF), Opaque Predicates, and their combination, and we measure both structural recovery of the control flow graph and preservation of program semantics. We evaluate five state-of-the-art large language models and show that CoT prompting significantly improves deobfuscation quality compared with simple prompting. We validate our approach on a diverse set of standard C benchmarks and report results using both structural metrics for control flow graphs and semantic metrics based on output similarity. Among the tested models and by applying CoT, GPT5 achieves the strongest overall performance, with an average gain of about 16\% in control-flow graph reconstruction and about 20.5\% in semantic preservation across our benchmarks compared to zero-shot prompting. Our results also show that model performance depends not only on the obfuscation level and the chosen obfuscator but also on the intrinsic complexity of the original control flow graph. Collectively, these findings suggest that CoT-guided large language models can serve as effective assistants for code deobfuscation, providing improved code explainability, more faithful control flow graph reconstruction, and better preservation of program behavior while potentially reducing the manual effort needed for reverse engineering.

\end{abstract}


\section{Introduction}

Code deobfuscation remains a significant challenge when using traditional analysis tools, such as static disassemblers and dynamic debuggers \cite{schrittwieser2016protecting,linn2003obfuscation}. These conventional methodologies often require extensive manual effort, specialized skills, and iterative analysis processes, greatly inflating the cost and complexity of reverse engineering tasks \cite{manuel2024enhancing}. Obfuscation techniques, such as Control Flow Flattening (CFF) and Opaque Predicates, further exacerbate these challenges by introducing substantial complexity into the compiled code. 

Obfuscation techniques disrupt logical execution paths and significantly complicate static analysis \cite{mohseni2025can}. Conventional static analyzers, such as disassemblers and decompilers, rely on structured and predictable code patterns \cite{novak2010taxonomy}; obfuscation intentionally breaks these patterns, making it difficult to reconstruct meaningful Control Flow Graphs (CFGs) and accurately interpret instructions \cite{balakrishnan2005code,xu2020layered}. Traditional reverse engineering tools such as \texttt{IDA Pro} and \texttt{Ghidra}, as well as standard debuggers such as \texttt{GDB} and \texttt{x64dbg}, offer limited capabilities for automating obfuscation analysis. Although useful for static and dynamic code analysis, these tools do not reliably identify and unravel widely used obfuscation methods, including CFF and opaque predicates. 

Traditional symbolic methods for code deobfuscation can work well on simple programs, but their cost grows rapidly as obfuscation makes the control flow and constraints more complex. A clear example is reported by Banescu et al., where a representative SAT instance takes about 7.5 seconds before obfuscation, but about 438 seconds after virtualization and control-flow flattening, resulting in a roughly 58-fold slowdown \cite{BanescuEtAl2017}. This result shows that the practical cost of deobfuscation is often measured by the total attack runtime, because solver time (such as SAT, SMT) is the main expense in automated analysis. The main reason for this high cost is that symbolic execution suffers from path explosion, which means the number of possible execution paths can grow very rapidly, increasing both time and memory use \cite{CadarSen2013,BaldoniEtAl2018}. In addition, symbolic memory operations and complex constraints can yield formulas that are too large or too complex for solvers to process efficiently. In stronger protection settings, this cost may shift from a delay of minutes to a point where the analysis becomes impractical, highlighting an important limitation of traditional symbolic tools for heavily obfuscated code \cite{XuEtAl2018}.

In the context of contemporary reverse engineering tools that integrate artificial intelligence techniques \cite{lachaux2021dobf}, Large Language Models (LLMs) have shown promise in high-level source code analysis \cite{fan2023static,sharma2021survey}. However, their effectiveness for code explainability and semantic analysis in low-level programming languages, such as assembly or Intermediate Representation, like LLVM-IR, remains underexplored. Furthermore, code obfuscation techniques continuously evolve, and LLMs struggle against novel obfuscation methodologies \cite{mohseni2025can}. These models require regular, potentially costly retraining or updates to maintain resilience against emerging obfuscation strategies, highlighting ongoing maintenance and adaptability as critical considerations for operational deployment. Beyond these challenges, we identified two practical barriers. First, training or fine-tuning an LLM specifically for code deobfuscation requires significant computational resources; yet, no evidence suggests that custom models outperform advanced commercial models like Claude Sonnet 4.5 or GPT-4o in code explainability tasks. Second, no comprehensive open-source dataset exists for training LLMs on automated code deobfuscation with explanatory annotations.

\begin{figure*}[h!]
  \centering  
  \includegraphics[width=0.85\textwidth]{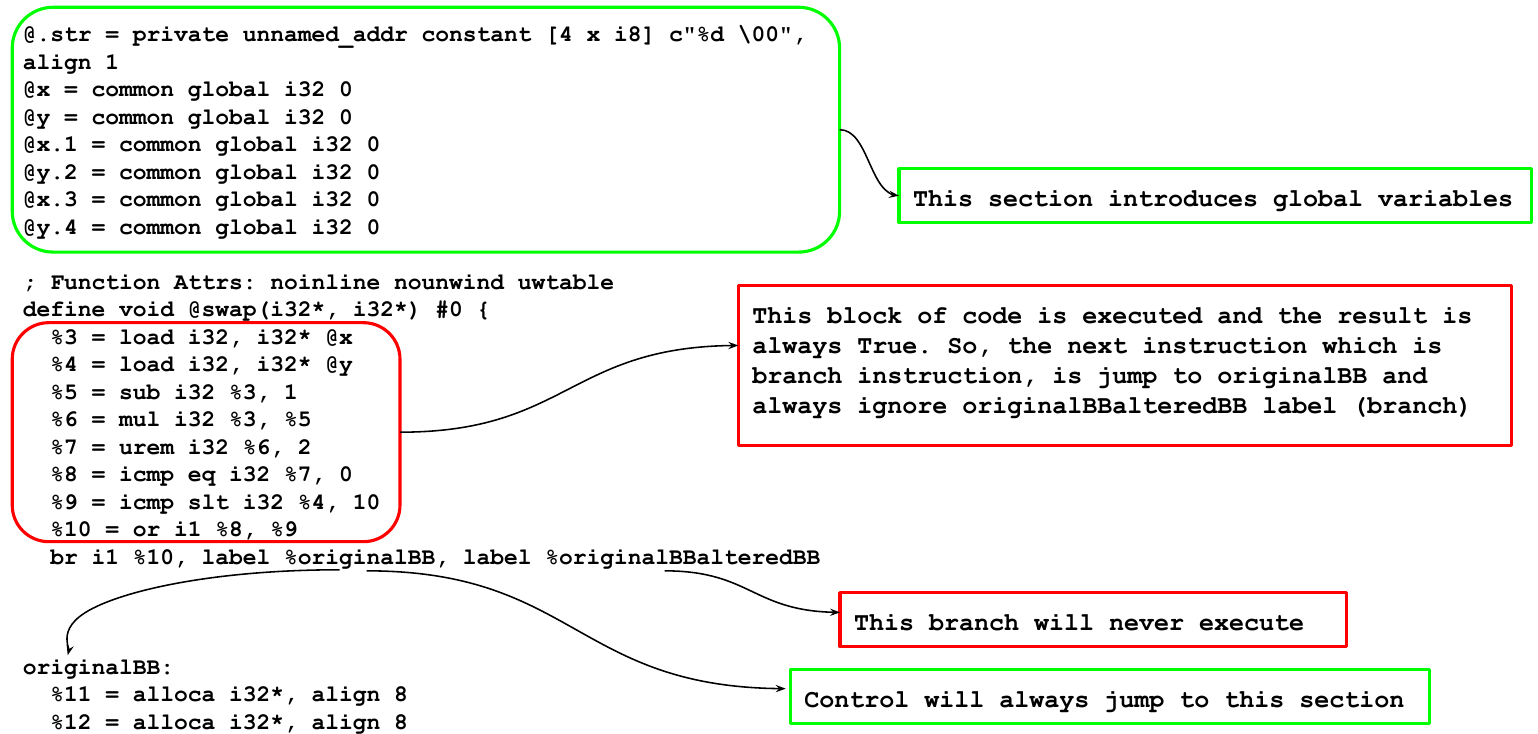}  
  \caption{The predicate in the highlighted region (red) is invariant and evaluates to \textbf{true} for all inputs; therefore, the conditional branch always targets \textit{originalBB}, rendering \textit{originalBBalteredBB} unreachable}
  \label{figure1}
\end{figure*}

To address these challenges, we leverage (CoT) prompting \cite{wei2022chain,sahoo2024systematic, zhang2022automatic, yu2023towards} with current state-of-the-art LLMs. CoT prompting offers a promising counter-strategy by enabling LLMs to decompose complex deobfuscation tasks into intermediate reasoning steps, systematically unraveling the layered transformations introduced by obfuscation. Through carefully designed prompts, we guide the model to trace execution paths, identify invariant properties, and distinguish genuine control flow from obfuscation artifacts. This structured reasoning approach allows LLMs to recognize patterns associated with common obfuscation techniques and recover underlying program semantics. In this paper, we focus exclusively on open-source obfuscators Tigress and O-LLVM \cite{junod2015obfuscator} for applying opaque predicates, CFF, and their combination.

In this paper, we make the following contributions:
\begin{itemize}[leftmargin=*, itemsep=0pt]
\item We demonstrate the capabilities of state-of-the-art LLMs to deobfuscate both low-level and high-level code, including LLVM-IR and C source code. Our approach combines CoT and procedural prompting with global variable tracing to enhance deobfuscation effectiveness.
\item We evaluate how different obfuscation techniques, such as opaque predicates, CFF, and their combination (Opaque-CFF), affect LLM deobfuscation performance across multiple models.
\item We systematically analyze the performance of selected LLMs using CoT prompting on obfuscated code samples generated by O-LLVM and Tigress.
\end{itemize}

\section{Background and Definitions}

\textbf{Code deobfuscation} reverses transformations applied to source code, assembly, or binary files that deliberately obscure program logic and structure, with the goal of restoring readability and interpretability \cite{udupa2005deobfuscation}. This process is critical in malware analysis \cite{chen2021impact}, particularly for metamorphic malware \cite{rad2012camouflage}, where obfuscation conceals malicious behavior from static analysis tools \cite{chandran2025static,asghar2024use}. Deobfuscation is also essential for intellectual property verification, enabling accurate assessment of code origins and licensing compliance.

\begin{figure*}[t]
  \centering  
  \includegraphics[width=0.85\textwidth]{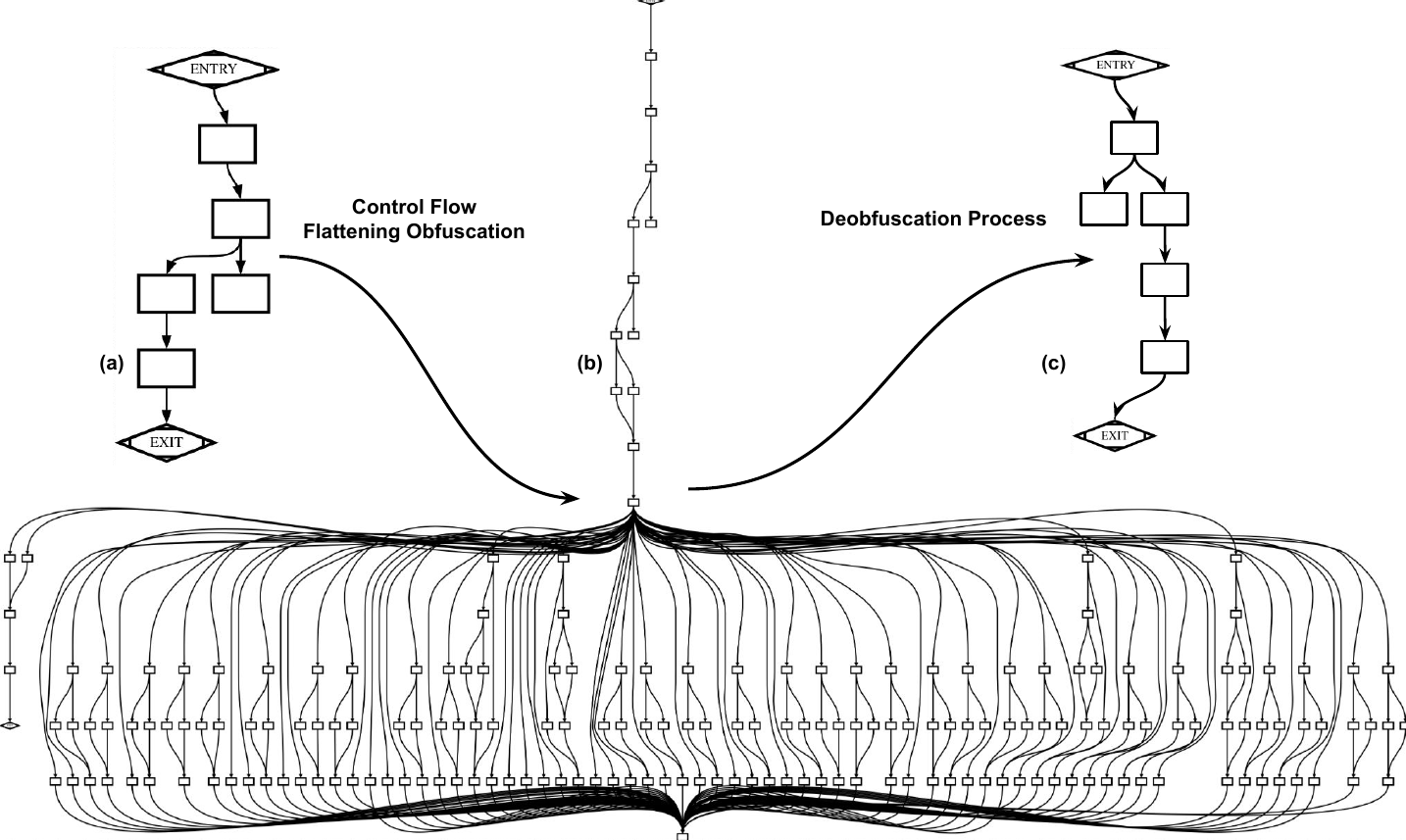}  
  \caption{Control Flow Graph (CFG) of \textbf{(a)} original code, \textbf{(b)} Obfuscated code by Tigress with Control Flow Flattening (CFF), and \textbf{(c)} Deobfuscated code by DeepSeek-V2, which shows some sort of hallucination.
  }
  \label{figure2}
\end{figure*}

\subsection{Layered Obfuscation Techniques}

Modern obfuscation strategies employ multiple techniques that can be applied independently or in combination, creating layered defenses against reverse engineering. We focus on two primary techniques, Opaque predicates and Control Flow Flattening (CFF), which can be deployed separately or combined to create more resilient obfuscation. Each layer adds computational complexity, opaque predicates introduce false execution paths, CFF obscures the program's logical structure, and their combination (Opaque-CFF) compounds both challenges simultaneously.

\noindent \textbf{Opaque Predicates.} An opaque predicate introduces conditional branches whose outcomes are predetermined but computationally difficult for static analysis to determine \cite{balachandran2016control,jin2024framework}. This technique creates bogus control flow (BCF) by adding unreachable branches that appear legitimate during analysis. Formally, a predicate $p$ with branch outcome $\tau$ is opaque if:

$$\exists \tau \in \{\top, \bot\}: \forall \mathbf{x} \in \text{domain}(p),\; p(\mathbf{x}) = \tau$$

where $\mathbf{x}$ represents program inputs, $\top$ denotes ``always true,'' and $\bot$ denotes ``always false.'' Regardless of input values, control flow consistently follows branch $\tau$, rendering alternative branches unreachable. The obfuscator typically inserts a new basic block before genuine code, with a predicate that conditionally jumps either to the real block or to a bogus block containing junk instructions. Common patterns exploit algebraic invariants such as $x^2 \geq 0$ or $x(x+1) \equiv 0 \pmod{2}$, which always evaluate to true but resist static analysis because automated tools struggle to prove mathematical invariants without expensive symbolic reasoning.

\noindent \textbf{Control Flow Flattening (CFF):} CFF restructures a program by decomposing it into basic blocks and routing execution through a centralized dispatcher, typically implemented as a loop with a switch statement \cite{johansson2017lightweight}. Let $\mathcal{C}$ denote the source code, $\mathcal{S}$ its Abstract Syntax Tree (AST), and $G_{cff}=(V,E)$ the state transition graph, where $V$ represents dispatcher states and $E \subseteq V \times V \times \Phi$ denotes conditional transitions guarded by predicates $\phi \in \Phi$. The semantic function $\llbracket \cdot \rrbracket: \mathcal{C} \times \mathcal{X} \to \mathcal{Y}$ maps code and inputs to outputs, preserving program behavior while obfuscating control flow structure. Figure \ref{figure2} illustrates how CFF transforms a simple control flow graph into a flattened structure and shows the deobfuscated result produced by DeepSeek-V2.

\noindent \textbf{Combined Layered Obfuscation (Opaque-CFF):} Applying both techniques in combination, embedding opaque predicates within control flow flattening, creates a layered obfuscation that is significantly more complex than either method alone. This multi-layer approach compounds the difficulty of both structural and semantic analysis, as analysts must simultaneously untangle the flattened dispatcher logic while identifying and removing fake branches introduced by opaque predicates.

\subsection{Deobfuscation via Global Variable Analysis}

Control flow deobfuscation is fundamentally a problem of variable provenance. Recovering the semantic roles of global variables, tracing their consumption points across code blocks, and determining how they shape branching conditions. When a model or human analyst can follow these variable interactions end-to-end, it becomes possible to distinguish genuine program logic from the artifacts of obfuscation, including the opaque predicates and spurious branches that CFF systematically introduces. Opaque predicates are particularly revealing targets of this analysis, because their defining characteristic is a branch condition whose truth value is fixed at construction time yet syntactically indistinguishable from a legitimate runtime decision; careful propagation of global variable states exposes this invariance. Figure~\ref{figure1} illustrates the process concretely in LLVM-IR. The upper portion of the listing declares the relevant global variables, while the lower portion exhibits an opaque predicate that unconditionally evaluates to true, permanently directing control flow to the genuine successor block and rendering the bogus alternative permanently unreachable. Identifying this pattern, a predicate whose value is structurally determined before execution, is precisely the kind of semantic inference that global variable analysis makes tractable.

\subsection{CoT Prompting for Deobfuscation}

Traditional deobfuscation tools struggle with layered obfuscation because they lack systematic reasoning strategies capable of decomposing compound transformations \cite{yadegari2015generic, bichsel2016statistical}.

Fine-tuning large language models on obfuscated code does not resolve this deficiency and introduces a distinct set of problems. Control flow flattening, for instance, can expand a compact function into an extensive state-driven dispatch structure, rapidly saturating a model's context window and forcing the analyst to segment the code in ways that obscure cross-block dependencies and the integrity of the control flow graph \cite{laszlo2009obfuscating, junod2015obfuscator}. Beyond the context limitation, fine-tuning on obfuscated corpora tends to reward surface pattern recognition rather than genuine semantic reasoning. A model may reliably eliminate opaque predicates that recur across training examples yet fail entirely on structurally novel predicates because it has internalized a lexical heuristic rather than learned to perform constraint propagation \cite{xu2017neural, ragkhitwetsagul2022toxic}. 
These limitations motivate a prompting-based approach. By eliciting explicit, step-by-step reasoning through CoT prompting \cite{wei2022chain}, one can guide a model to externalize its deobfuscation process, enforce symbolic discipline at each inferential step, and maintain coherent tracking of global variable states and control flow semantics without any modification to model weights.

Concretely, CoT prompting operationalizes this reasoning scaffold by guiding the model through a structured sequence of inferential steps: (a) parsing the abstract syntax tree to locate dispatcher structures, (b) tracing state variables across basic blocks to establish their provenance and (c) consumption patterns, evaluating branch predicates for value invariance, and reconstructing the original control flow graph by excising dead branches whose reachability is structurally precluded. Each step externalizes an element of the reasoning chain that would otherwise remain implicit in a single-pass generation, making the model's intermediate commitments inspectable and correctable.

This decomposition is particularly well suited to control flow deobfuscation because it mirrors, at scale and with reproducible consistency, the logical procedure that a skilled reverse engineer performs manually. Tracking global variable states, isolating algebraic invariants embedded in opaque predicates, and systematically dismantling dispatcher artifacts are not independent heuristics but stages of a unified semantic recovery process. By encoding this process explicitly in the prompt, CoT prompting preserves program semantics throughout deobfuscation rather than approximating a plausible-looking output, which is precisely the discipline that token-level generation alone cannot guarantee.

\section{Related Work}

\textit{LLM supporting deobfuscation}: Recent studies show that large language models can support code deobfuscation, but results remain mixed and depend on the setting. Patsakis et al.\cite{patsakis2024assessing} tested four LLMs on obfuscated PowerShell scripts taken from real malware campaigns. They found that GPT-4 produced the strongest results among the tested models, but the study also highlighted important limitations, the models sometimes produced false outputs or refused to answer. Beste et al. \cite{beste2025exploring} reported that fine-tuned code models can outperform GPT-4 and compiler-based baselines on obfuscated C code, although preserving the original program behavior becomes more difficult as the obfuscation grows more complex. In a related study, Feng et al. \cite{feng2025can} found that GPT-4.1-mini gave the strongest overall results and that KLEE-based artifacts improved both compilation success and behavior preservation when the models were trained to use them.

\textit{NLP representation of code}: Recent research shows that machine learning can help recover the meaning of hidden code by treating programs as a form of language. In this view, a program is studied in a manner similar to how a sentence is studied in natural language processing. Tsfaty et al. (2023) \cite{tsfaty2023malicious} showed that deep learning models originally developed for text translation can map obfuscated code back to a clearer, more readable form. This idea is useful because it learns patterns directly from data rather than relying solely on fixed rules written by experts. Ndichu et al. (2020) \cite{ndichu2020deobfuscation} further showed that such methods can clean the code and remove confusing parts before deeper analysis begins. Tayyab et al. (2022)\cite{tayyab2022survey} also explained that deep learning can automatically extract important software features, reducing manual work and the need for traditional rule-based methods.

\textit{Structure of code}: Another important research direction focuses on the structure of code rather than only on its text. Tang et al. (2022) \cite{tang2022dfsgraph} addressed the control flow flattening problem with graph neural networks that model how data flows through the program, helping the model look past the confusing surface structure. Chen et al. (2021) \cite{chen2021deep} and Hassan et al. (2021) \cite{hassan2021neural} showed that deep learning models can predict the operations needed to recover the intended logic even when the structure has been heavily changed. Jain et al. (2024) \cite{jain2024towards} further found that training on generated data helps these structural models restore unknown values more safely and accurately. These results suggest that structure-based learning is especially useful when the main challenge comes from altered control paths or hidden data relationships.

Although LLMs support deobfuscation, NLP, and code structure, prior work suffers from two key limitations. First, they add more complexity to the deobfuscation process. Second, they have not been sufficiently evaluated in the layered obfuscation setting. We address these gaps by applying CoT prompting with off-the-shelf reasoning models to statically deobfuscate LLVM-IR and C code under combined obfuscation layers (Opaque-CFF), systematically evaluating CFG reconstruction without requiring model training, dynamic execution, or language-specific runtime analysis.

\section{Methodology}

\autoref{alg} presents the conceptual framework underlying our CoT prompting strategy for code deobfuscation. Rather than an executable procedure, this algorithm represents the logical reasoning structure that CoT prompts guide the LLM to follow. The framework comprises five phases that systematically identify and remove obfuscation constructs.

\textbf{Phase 1: Obfuscation Detection} (lines 1-4). The model first analyzes the code structure to identify obfuscation patterns. By parsing the Abstract Syntax Tree (AST) representation, the model detects dispatcher constructs characteristic of CFF (line 2), extracts state variables $\sigma$ that control dispatcher transitions (line 3), and identifies opaque predicate patterns $\mathcal{P}_{op}$ using algebraic invariant recognition (line 4). This phase determines whether deobfuscation is necessary and which techniques (Opaque, CFF, or both) are present.

\textbf{Phase 2: Control Flow Flattening Recovery} (lines 5-15). For code exhibiting CFF, the model extracts individual case blocks $\mathcal{B} = \{B_1, B_2, \ldots, B_n\}$ from the dispatcher switch statement (line 5) and constructs a state transition graph $G_{cff} = (V, E)$ by analyzing how state variable $\sigma$ evolves across blocks. For each basic block $B_i$, the model identifies its current state label $s_{curr}$ (line 8), analyzes transitions to determine successor states and their guard conditions $(s_{next}, \phi)$ (line 9), and builds the directed graph by adding state vertices and conditional edges (lines 10-12). The initial state $s_0$ is identified by tracing $\sigma$'s initialization (line 15).

\textbf{Phase 3: Control Flow Reconstruction} (lines 16-18). Using the recovered state transition graph, the model reconstructs the original control flow by detecting loop back-edges $\mathcal{L}$ (line 16), performing topological sorting to determine the natural execution order $\pi$ while respecting loop structures (line 17), and generating cleaned control flow graph $C_{cfg}$ by sequentially arranging basic blocks according to $\pi$ (line 18).

\textbf{Phase 4: Opaque Predicate Elimination} (lines 19-26). The model evaluates each identified opaque predicate $p \in \mathcal{P}_{op}$ to determine its invariant outcome $\tau(p)$ (line 20). For predicates that always evaluate to true ($\tau(p) = \top$), the model replaces the branching construct with only the then-branch (line 22); for those always false ($\tau(p) = \bot$), only the else-branch is retained (line 24). This eliminates unreachable bogus branches introduced during obfuscation.

\textbf{Phase 5: Code Cleanup} (lines 27-29). The model identifies dead variables $\mathcal{V}_{dead}$, including the dispatcher state variable $\sigma$ and any variables with zero reference counts (line 27), and removes all dead code and unused variables (line 28). Finally, \textsc{NormalizeAST} applies syntactic normalization including consistent formatting, expression simplification, and structural cleanup to produce the final deobfuscated code $C_{deobf}$ (line 29).

Throughout this process, CoT prompting provides explicit reasoning scaffolding at each phase, guiding the model to articulate intermediate steps such as ``this predicate evaluates to true because $x^2 \geq 0$ for all $x$'' or ``state variable transitions from $s_3$ to $s_7$ when condition $\phi$ holds.'' This structured reasoning enables the model to systematically decompose complex obfuscation patterns that would be difficult to reverse through pure pattern matching.

\algnewcommand\algorithmicinput{\textbf{Input:}}
\algnewcommand\Input{\item[\algorithmicinput]}
\algnewcommand\algorithmicoutput{\textbf{Output:}}
\algnewcommand\Output{\item[\algorithmicoutput]}

\begin{algorithm}[t]
\caption{Code Deobfuscation with CoT}
\label{alg}
\begin{algorithmic}[1]
\Input Obfuscated source code $\mathcal{C}_{obf}$
\Output Deobfuscated source code $\mathcal{C}_{deobf}$

\Statex \textit{/* Phase 1: Obfuscation Detection */}
\State $\mathcal{S} \gets \textsc{ParseAST}(\mathcal{C}_{obf})$ 
\State $\mathcal{D} \gets \textsc{DetectDispatcher}(\mathcal{S})$
\State $\sigma \gets \textsc{ExtractStateVar}(\mathcal{D})$
\State $\mathcal{P}_{op} \gets \textsc{IdentifyOpaquePredicates}(\mathcal{S})$

\Statex \textit{/* Phase 2: Control Flow Flattening Recovery */}
\State $\mathcal{B} \gets \{B_1, B_2, \ldots, B_n\} \gets \textsc{Ex-Cases}(\mathcal{D})$
\State $G_{cff} = (V, E) \gets \emptyset$
\ForAll{$B_i \in \mathcal{B}$}
    \State $s_{curr} \gets \textsc{GetCaseLabel}(B_i)$
    \State $\mathcal{T}_i \gets \textsc{AnalyzeTransitions}(B_i, \sigma)$
    \ForAll{$(s_{next}, \phi) \in \mathcal{T}_i$}
        \State $V \gets V \cup \{s_{curr}, s_{next}\}$
        \State $E \gets E \cup \{(s_{curr}, s_{next}, \phi)\}$
    \EndFor
\EndFor
\State $s_0 \gets \textsc{GetInitialState}(\sigma)$

\Statex \textit{/* Phase 3: Control Flow Reconstruction */}
\State $\mathcal{L} \gets \textsc{DetectBackEdges}(G_{cff})$
\State $\pi \gets \textsc{TopologicalSort}(G_{cff}, s_0, \mathcal{L})$ 
\State $\mathcal{C}_{cfg} \gets \textsc{ReconstructCFG}(\pi, \mathcal{B}, \mathcal{L})$

\Statex \textit{/* Phase 4: Opaque Predicate Elimination */}
\ForAll{$p \in \mathcal{P}_{op}$}
    \State $\tau(p) \gets \textsc{EvaluatePredicate}(p)$
    \If{$\tau(p) = \top$}
        \State $\mathcal{C}_{cfg} \gets \textsc{ReplaceWith}(\mathcal{C}_{cfg}, p, p.\textit{then\_branch})$
    \ElsIf{$\tau(p) = \bot$}
        \State $\mathcal{C}_{cfg} \gets \textsc{ReplaceWith}(\mathcal{C}_{cfg}, p, p.\textit{else\_branch})$
    \EndIf
\EndFor

\Statex \textit{/* Phase 5: Code Cleanup */}
\State $\mathcal{V}_{dead} \gets \{v \mid v \in \textsc{Vars}(\mathcal{C}_{cfg}) \land \textsc{RefCount}(v) = 0\}$
\State $\mathcal{C}_{clean} \gets \textsc{RemoveDeadCode}(\mathcal{C}_{cfg}, \mathcal{V}_{dead} \cup \{\sigma\})$
\State $\mathcal{C}_{deobf} \gets \textsc{NormalizeAST}(\mathcal{C}_{clean})$

\State \Return $\mathcal{C}_{deobf}$
\end{algorithmic}
\end{algorithm}


\begin{table*}[t]
\centering
\caption{Evaluation of Models for Deobfuscation (LLVM Obfuscator). CoT rows include relative improvement over zero-shot.}
\label{table1}
\footnotesize
\resizebox{\textwidth}{!}{
\begin{tabular}{lcccccc}
\toprule
 & \multicolumn{2}{c}{Opaque} & \multicolumn{2}{c}{CFF} & \multicolumn{2}{c}{Opaque-CFF} \\
\cmidrule(lr){2-3} \cmidrule(lr){4-5} \cmidrule(lr){6-7}
Model & SRS\% & BLEU\% & SRS\% & BLEU\% & SRS\% & BLEU\% \\
\midrule

GPT5 (Zero shot) & 86.2 & 77.5 & 87.0 & 85.6 & 83.6 & 80.2 \\
GPT5 with CoT 
& \textbf{99.6 (+15.5\% $\uparrow$)} 
& \textbf{100 (+29.0\% $\uparrow$)} 
& \textbf{99.7 (+14.6\% $\uparrow$)} 
& \textbf{98.5 (+15.1\% $\uparrow$)} 
& \textbf{100 (+19.6\% $\uparrow$)} 
& \textbf{99.0 (+23.4\% $\uparrow$)} \\

\midrule

o3 (Zero shot) & 57.2 & 42.7 & 51.6 & 41.0 & 35.8 & N/A \\
o3 with CoT 
& \textbf{79.2 (+38.5\% $\uparrow$)} 
& \textbf{76.7 (+79.6\% $\uparrow$)} 
& \textbf{83.9 (+62.6\% $\uparrow$)} 
& \textbf{80.2 (+95.6\% $\uparrow$)} 
& \textbf{74.3 (+107.5\% $\uparrow$)} 
& N/A \\

\midrule

DeepSeek-V2 (Zero shot) & 84.2 & 71.0 & 85.3 & 73.6 & 82.4 & 79.5 \\
DeepSeek-V2 with CoT 
& \textbf{95.4 (+13.3\% $\uparrow$)} 
& \textbf{97.2 (+36.9\% $\uparrow$)} 
& \textbf{93.7 (+9.8\% $\uparrow$)} 
& \textbf{96.5 (+31.1\% $\uparrow$)} 
& \textbf{89.1 (+8.1\% $\uparrow$)} 
& \textbf{92.3 (+16.1\% $\uparrow$)} \\

\midrule

Qwen-3 Max (Zero shot) & 78.3 & 75.9 & 84.1 & 71.0 & 77.5 & 81.3 \\
Qwen-3 Max with CoT 
& \textbf{92.2 (+17.8\% $\uparrow$)} 
& \textbf{94.7 (+24.8\% $\uparrow$)} 
& \textbf{89.0 (+5.8\% $\uparrow$)} 
& \textbf{93.3 (+31.4\% $\uparrow$)} 
& \textbf{83.2 (+7.4\% $\uparrow$)} 
& \textbf{91.4 (+12.4\% $\uparrow$)} \\

\midrule

QWQ-32B (Zero shot) & 64.5 & 47.9 & 61.6 & 53.0 & 46.1 & N/A \\
QWQ-32B with CoT 
& \textbf{85.3 (+32.2\% $\uparrow$)} 
& \textbf{86.5 (+80.6\% $\uparrow$)} 
& \textbf{81.0 (+31.5\% $\uparrow$)} 
& \textbf{87.0 (+64.2\% $\uparrow$)} 
& \textbf{73.8 (+60.1\% $\uparrow$)} 
& \textbf{84.6 (N/A)} \\

\bottomrule
\end{tabular}}
\end{table*}

\begin{table*}[t]
\centering
\caption{Evaluation of Models for Deobfuscation (Tigress Obfuscator). CoT rows include relative improvement over zero-shot.}
\label{table2}
\footnotesize
\resizebox{\textwidth}{!}{
\begin{tabular}{lcccccc}
\toprule
 & \multicolumn{2}{c}{Opaque} & \multicolumn{2}{c}{CFF} & \multicolumn{2}{c}{Opaque-CFF} \\
\cmidrule(lr){2-3} \cmidrule(lr){4-5} \cmidrule(lr){6-7}
Model & SRS\% & BLEU\% & SRS\% & BLEU\% & SRS\% & BLEU\% \\
\midrule

GPT5 (Zero shot) & 81.3 & 69.7 & 84.0 & 80.0 & 81.6 & 76.8 \\
GPT5 with CoT 
& \textbf{100 (+23.0\% $\uparrow$)} 
& \textbf{99.0 (+42.0\% $\uparrow$)} 
& \textbf{100 (+19.0\% $\uparrow$)} 
& \textbf{98.1 (+22.6\% $\uparrow$)} 
& \textbf{100 (+22.5\% $\uparrow$)} 
& \textbf{98.6 (+28.4\% $\uparrow$)} \\

\midrule

o3 (Zero shot) & 51.1 & 32.0 & N/A & 42.1 & 21.8 & N/A \\
o3 with CoT 
& \textbf{75.2 (+47.2\% $\uparrow$)} 
& \textbf{72.4 (+126.3\% $\uparrow$)} 
& \textbf{76.7 (N/A)} 
& \textbf{74.8 (+77.7\% $\uparrow$)} 
& \textbf{70.6 (+223.9\% $\uparrow$)} 
& \textbf{69.0 (N/A)} \\

\midrule

DeepSeek-V2 (Zero shot) & 77.5 & 68.0 & 54.5 & 59.6 & 51.2 & 61.4 \\
DeepSeek-V2 with CoT 
& \textbf{91.6 (+18.2\% $\uparrow$)} 
& \textbf{94.3 (+38.7\% $\uparrow$)} 
& \textbf{88.4 (+62.2\% $\uparrow$)} 
& \textbf{94.9 (+59.2\% $\uparrow$)} 
& \textbf{85.8 (+67.6\% $\uparrow$)} 
& \textbf{90.4 (+47.2\% $\uparrow$)} \\

\midrule

Qwen-3 Max (Zero shot) & 67.9 & 60.4 & 57.0 & 52.9 & 48.3 & 53.6 \\
Qwen-3 Max with CoT 
& \textbf{90.2 (+32.9\% $\uparrow$)} 
& \textbf{90.1 (+49.2\% $\uparrow$)} 
& \textbf{86.0 (+50.9\% $\uparrow$)} 
& \textbf{92.4 (+74.7\% $\uparrow$)} 
& \textbf{82.1 (+69.9\% $\uparrow$)} 
& \textbf{87.3 (+62.9\% $\uparrow$)} \\

\midrule

QWQ-32B (Zero shot) & 54.0 & 38.2 & 59.4 & 35.1 & N/A & N/A \\
QWQ-32B with CoT 
& \textbf{74.9 (+38.7\% $\uparrow$)} 
& \textbf{76.7 (+100.8\% $\uparrow$)} 
& \textbf{73.2 (+23.2\% $\uparrow$)} 
& \textbf{80.0 (+127.9\% $\uparrow$)} 
& \textbf{70.6 (N/A)} 
& \textbf{79.1 (N/A)} \\

\bottomrule
\end{tabular}}
\end{table*}

\subsection{Models and Specification}
When dealing with obfuscated code exhibiting these complex transformations, we observe that reasoning models perform substantially better than standard language models. Control flow deobfuscation requires following intricate code paths, maintaining state across multiple basic blocks, and tracking how control flow changes during execution tasks that demand extended CoT inference. Standard LLMs, when confronted with heavily obfuscated programs, often produce only local edits or generate code that fails to compile. By contrast, reasoning models such as GPT5 and DeepSeek-V2 can propose step-by-step changes that restore original program structure while preserving semantic behavior, explain intermediate reasoning transparently, revise draft answers across refinement rounds, and correct logical errors systematically. This capacity for explicit, verifiable reasoning directly addresses the requirements imposed by Algorithm \autoref{alg}'s structured pipeline.

To study this potential, we evaluate several state-of-the-art reasoning models, GPT5 and o3 from OpenAI, Qwen-3 MAX and QwQ-32B from Qwen, and DeepSeek-V2. Our goal is to measure how effectively these models transform control-flow obfuscated code (under Opaque, CFF, and Opaque-CFF configurations) into semantically equivalent, readable implementations while accurately reconstructing the control flow graphs. For all models, we use a 128k token context window and allow up to 100k output tokens. Temperature is set to 0.8 except for o3, which uses OpenAI's internal settings. Each model operates in reasoning mode to enable extended CoT inference during deobfuscation.

\subsection{Evaluation}

To ensure a comprehensive assessment, we use two complementary metrics that capture both structural and functional improvements resulting from deobfuscation. Given that the deobfuscation process primarily involves the elimination of opaque predicates and the removal of unreachable or redundant control flow branches, one of the most immediate and measurable effects is the \textbf{Structural Similarity Score (SRS)} \cite{li2019graph, boobalan2016graph}. The SRS is a metric that measures how closely two Control Flow Graphs (CFGs) are structured. Let $G_1 = (V_1, E_1)$ be the CFG of the original code and $G_2 = (V_2, E_2)$ be the CFG of the deobfuscated code. Let $\mathrm{GED}(G_1,G_2)$ be the Graph Edit Distance (GED) \cite{ chan2014method}, which is the minimum number of node and edge edits needed to turn $G_1$ into $G_2$. We first define the normalized distance:

$$d(G_1,G_2) = \frac{\mathrm{GED}(G_1,G_2)}{\max\left(|V_1| + |E_1|,\ |V_2| + |E_2|\right)}$$

Then, the Structural Similarity Score is defined as follows:

$$\mathrm{SRS}(G_1,G_2) = 1 - d(G_1,G_2)$$

This means that SRS is always between 0 and 1, with 1 indicating that the graphs are structurally identical and 0 indicating that they are very different. SRS is preferred for measuring similarity between CFGs because it transforms a raw edit cost into a clear and normalized similarity value, which makes it easier to compare different deobfuscation results in a fair and interpretable manner.

One important factor in the deobfuscation process is preserving the code's semantics. In order to measure the semantic and final performance of two sets of code (both the original code and the deobfuscated code), we apply the \textbf{BLEU (Bilingual Evaluation Understudy)} score \cite{papineni2002bleu, marasek2015enhanced, post2018call} to the outputs of both programs. The BLEU score is a metric that measures the similarity between two pieces of text by comparing their word patterns. A score close to 1 means that the candidate text is very similar to the reference text, while a score near 0 indicates that they are quite different. We use BLEU as a simple, reproducible, and lightweight lexical baseline because it measures token-level n-gram overlap and helps estimate the surface similarity between recovered code and reference code.
    
\begin{figure}[t]
        \centering
        \includegraphics[width=0.75\columnwidth]{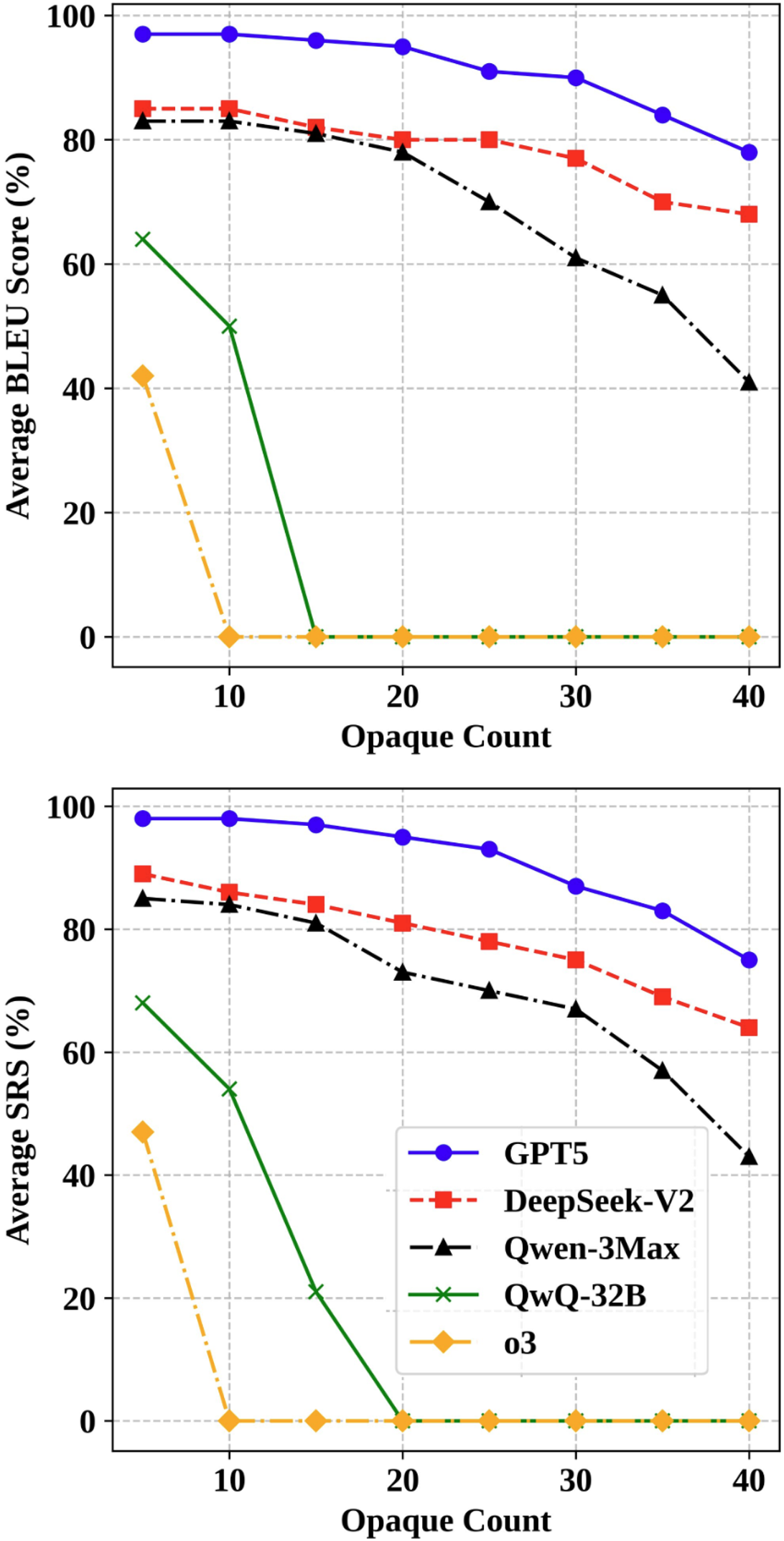}
        \caption{BLEU and SRS scores for \textit{Group 2} experiment 2.}
        \label{fig:group2}
\end{figure}
    
BLEU captures both local wording and the structure of the output. When the deobfuscated program produces text that uses the same words and short phrases as the original, in a similar order and with a similar length, BLEU scores are high. This suggests that the program still performs the same task and returns results very close to the original, even if the internal code has been transformed. If the deobfuscation process fails and the program's behavior changes, the output will differ, and BLEU will assign a lower score. We note that in real world malware analysis, exact replication of logic is not the default goal; the goal is to understand what is being done, assess risk, attribute techniques, and make various downstream determinations. Exact logic reconstruction is needed for some tasks, but imperfect logic reconstruction is still a valuable output.

\begin{figure}[t]
        \centering
        \includegraphics[width=0.75\columnwidth]{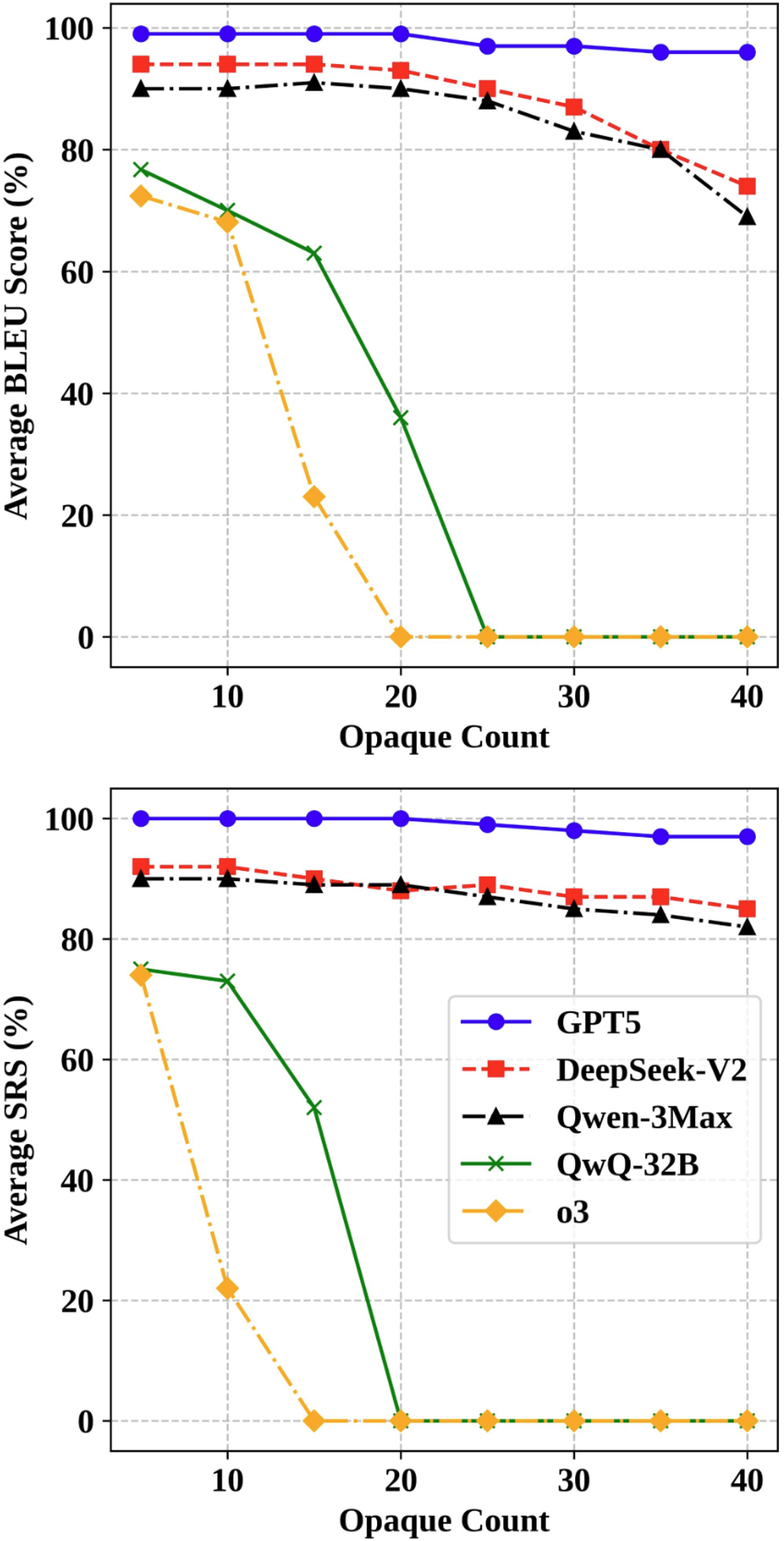}
        \caption{BLEU and SRS scores for \textit{Group 1} in experiment 2.}
        \label{fig:group1}
\end{figure}

\subsection{Codes and Experiments}
To run our experiments on code deobfuscation using CoT, we follow common practices in the literature and utilize a set of well-known computer science algorithms as benchmarks. All programs are written in standard C and span a range of control flow complexities; thus, they provide a diverse testbed for the models. All code is obfuscated using the open source tools Tigress and O-LLVM. Before being provided to the models, we remove all code names and comments so that the models cannot rely on high-level hints about the algorithm's type or structure.

\begin{figure}[t]
\centering
\begin{tikzpicture}
\begin{groupplot}[
    group style={
        group size=1 by 2,
        vertical sep=0.95cm
    },
    width=\columnwidth,
    height=0.15\textheight,
    ybar,
    ymin=0,
    ymax=105,
    symbolic x coords={GPT5,DeepSeekV2,Qwen3Max,QWQ32B,o3},
    xtick=data,
    enlarge x limits=0.06,
    ylabel near ticks,
    tick label style={font=\footnotesize},
    label style={font=\small\bfseries},
    xticklabel style={font=\footnotesize\bfseries},
    every axis plot/.append style={
        fill=blue!90,
        draw=blue!90
    },
    grid=major,
    grid style={dashed,gray!35},
    axis background/.style={fill=gray!8},
    axis line style={black},
    tick style={black}
]

\nextgroupplot[
    ylabel={Average BLEU(\%)}
]
\addplot coordinates {
    (GPT5,99)
    (DeepSeekV2,94)
    (Qwen3Max,91)
    (QWQ32B,82)
    (o3,76)
};

\nextgroupplot[
    ylabel={Average SRS(\%)}
]
\addplot coordinates {
    (GPT5,100)
    (DeepSeekV2,90)
    (Qwen3Max,87)
    (QWQ32B,76)
    (o3,77)
};

\end{groupplot}
\end{tikzpicture}
\caption{Average BLEU and SRS scores across models.}
\label{fig:bleu_srs_barplots}
\end{figure}
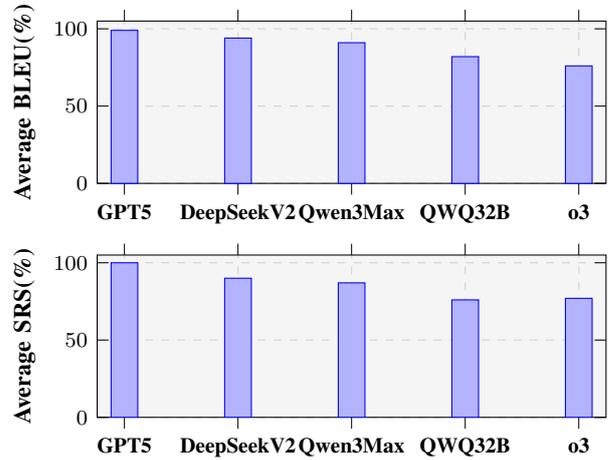

We evaluate our deobfuscation pipeline on a fixed pool of 12 standard C benchmark programs that cover a wide range of control-flow graph (CFG) complexity. The pool is divided into two groups, Group 1 contains nine lower-complexity programs (Merge Sort, Heap Sort, Quick Sort, Binary Search, Dijkstra, BFS, DFS, Knapsack, and Matrix Multiplication), and Group 2 contains three higher-complexity programs (N-Queens with solution printing, an AVL tree with rotations, and Huffman encoding/decoding). For every benchmark, we generate obfuscated variants using both Tigress and O-LLVM and keep this code pool constant across all model evaluations. Before prompting the models, we remove algorithm-identifying names and comments so that the models must rely on the code's executable structure rather than high-level hints. In Experiment 1, each benchmark is obfuscated under three conditions, Opaque Predicates only, CFF only, and Opaque + CFF so that each original program yields a consistent set of variants per obfuscator, enabling controlled comparison across different obfuscation modes.

To ensure reproducibility, we apply the same obfuscation configuration and experimental conditions across all models, and we report the exact tool settings used to generate each variant. For both Tigress and O-LLVM, we specify the exact tool versions used in the Debian Linux environment, the full command lines, including all flags for opaque predicates and control-flow flattening, and the randomness controls, such as fixed seeds or deterministic modes. We also compile all programs with GCC 4.6 on Linux Debian to keep the build environment consistent. To ensure a fair experimental setup, we used the compiler’s default switches and options, so the compiler did not introduce any additional noise during obfuscation or program execution In Experiment 2, we isolate the effect of opaque predicates by applying a single obfuscation layer and increasing the number of opaque branches from 0 to 40 in steps of 5, while keeping all other factors unchanged. This design allows us to study how model performance changes as opaque predicate density increases, under a tightly controlled, fully documented protocol.

\section{Results and Discussion}

To deobfuscate opaque branches, models must identify and eliminate opaque predicates and unreachable branches introduced during obfuscation. Deobfuscating CFF elements requires models to accurately identify dispatcher-switch constructs and eliminate all associated flattening components, including case statements linked to the dispatcher.

The experiment includes an ablation study comparing CoT prompting with non-CoT prompting to quantify the extent to which CoT contributes to deobfuscation. In this setting, we use the same benchmarks and models but change only the prompting strategy, and we then compare the results. \autoref{table1} and \autoref{table2} compare simple prompting and CoT prompting for deobfuscation, showing that CoT prompting significantly improves how well the models reconstruct the original code. In particular, the CoT setting yields much higher reconstruction quality and more accurate recovery of program structure, indicating that explicitly guiding the model through a step-by-step reasoning process is a key factor in successful LLM-based deobfuscation. This behavior shows that the CoT prompting strategy is not only effective when deobfuscation is required but is also careful and selective, as it can identify when deobfuscation is unnecessary and maintain the integrity of the input program.

For experiment 1, \autoref{table1} and \autoref{table2} summarize how all models perform when deobfuscating code that has been protected with opaque, CFF, and combined opaque–CFF transformations in both O-LLVM and Tigress for experiment 1. GPT5 provides the strongest results across all settings. Its high SRS values indicate that it almost fully rebuilds the original control flow graphs, and its BLEU scores show that the deobfuscated outputs are almost identical to the original program outputs at the semantic level. DeepSeek-V2 and Qwen-3 MAX achieve very similar scores to each other, with a small but consistent drop in CFF compared to opaque branches. Furthermore, DeepSeek-V2 is slightly better at restoring the original graph structure than Qwen. QwQ-32B recovers about 73.8\% of the original structure for O-LLVM and 70.6\% for Tigress under the combined opaque–CFF setting, which is close to the performance of the o3 model. Overall, all models achieve acceptable levels of structural and semantic recovery, but there are clear differences among obfuscators. O-LLVM is easier to reverse because its opaque predicates have limited variation, even under control-flow flattening, which restricts the search space for models, whereas Tigress introduces greater structural diversity. This is reflected in lower average scores and more frequent errors when the models attempt to reconstruct the deobfuscated code. \textit{N/A} indicates failure to produce a compilable program or failure to execute under the evaluation harness. 

Regarding experiment 2, figure \ref{fig:group1} provides clear evidence of the robustness of the models when the original program has a simple control flow graph (Group 1 with a simple CFG). GPT5 shows the highest resilience, its BLEU score remains almost constant as we add more opaque branches, and its high SRS score indicates that it can still reconstruct the original code's control flow graph with little loss of structure. DeepSeek-V2 and Qwen-3 MAX follow a similar pattern, but their BLEU and SRS values decrease steadily as the number of opaque predicate branches increases, which means that their deobfuscation quality declines as the noise grows. QwQ-32B shows slightly better resistance than these models, but its scores also decline, and its trend is close to that of o3. In contrast, o3 degrades much more quickly; its BLEU score is already heavily reduced at 15 opaque branches, and it fails to produce a valid deobfuscated program once the number of opaque predicates exceeds 20.

\begin{figure}[h!]
  \centering  
  \includegraphics[width=0.49\textwidth]{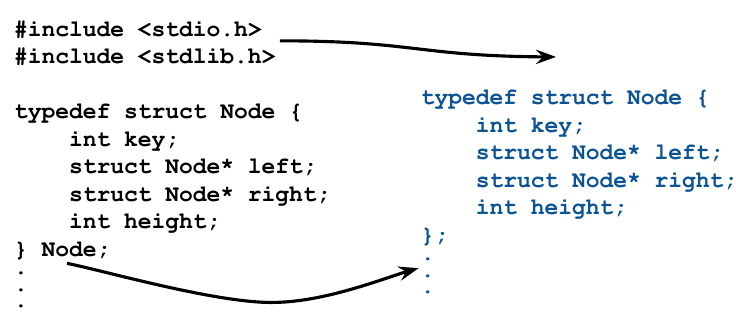}  
  \caption{\textbf{(Left)} Original snippet code. \textbf{(Right)} Deobfuscated snippet code by DeepSeek-V2. Headers and Node structure declaration are missing. Original code is obfuscated by Tigress.}
  \label{figure5}
\end{figure}

\begin{figure*}[t]
  \centering  
  \includegraphics[width=1\textwidth]{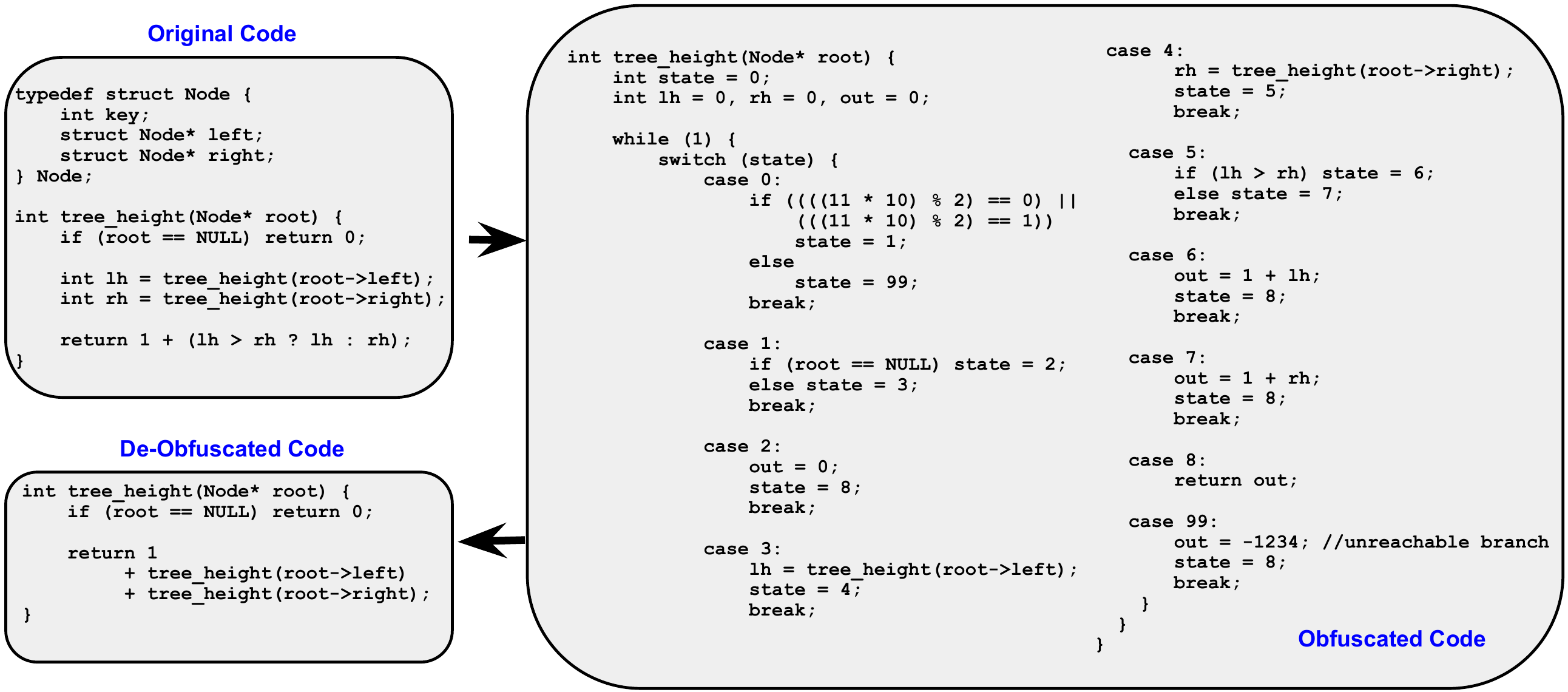}  
  \caption{Illustrative semantic hallucination during code deobfuscation. The original function computes the height of a binary tree. The hallucinated deobfuscated output is clean, short, and compilable, but it changes the program semantics by summing the heights of both subtrees instead of taking their maximum.}
  \label{figure6}
\end{figure*}

Figure \ref{fig:group2} shows a stronger decline across all metrics than figure \ref{fig:group1}, indicating that the models are more sensitive when the original program has a complex control flow graph (Group 2). For GPT5, the BLEU score remains almost flat until we reach 20 opaque branches, after which it decreases almost linearly, falling below 0.8 when the number of opaque branches reaches 40; the SRS curve follows the same pattern, indicating parallel loss in structural recovery. DeepSeek-V2 and the Qwen models also lose performance as we add more opaque predicates, but DeepSeek-V2 remains clearly better at producing correct deobfuscated code. QwQ-32B recovers roughly half of the correct output when there are 10 opaque branches; however, it fails to deobfuscate the program when additional opaque predicates are introduced. The o3 model can remove only a small number of opaque branches before it also fails. Taken together, these results show that, beyond the strength and design of the obfuscation tools and techniques, the inherent complexity of the original control flow graph directly and negatively affects the deobfuscation performance of both LLM and LRM models. Figure \ref{fig:bleu_srs_barplots} shows the overall average performance of models for retrieving BLEU and SRS scores.

One important observation from our study is that model performance improves when we explicitly show it the pattern of opaque predicates in the CoT prompt. Many open source and commercial obfuscation tools generate opaque predicates statically and reuse a small set of structural patterns to build opaque branches. By adding a few-shot learning segment to the prompt that shows examples of these typical patterns and explains why the branch condition is always true or always false, we provide the model with concrete clues about the nature of opaque branches. In our experiments, this guidance led to noticeable gains in both structural and semantic metrics, as the models were better able to identify and remove opaque branches while preserving the program's real control flow and behavior. This result suggests that even a small amount of targeted prior knowledge about common opaque predicate schemes can significantly improve the effectiveness of LLM-based deobfuscation.

\textbf{Hallucination:} Across our experiments, all models showed some level of hallucination; however, the type and severity of these errors differed between models and became worse as the obfuscation became stronger. In Experiment 2, when we increased the number of opaque branches, some models, such as QwQ-32B and o3, could no longer produce even a generic deobfuscated version of the program, suggesting they failed to recover the program's underlying structure. By contrast, as the number of opaque branches increased, DeepSeek-V2 and the Qwen models still generated code; however, they hallucinated more often and sometimes omitted important code elements, even while maintaining the main control flow structure of the original program. 

Based on our observation, hallucination in LLM-based deobfuscation can take several forms and is closely related to the strength and layering of the obfuscation, as well as the complexity of the original control flow graph. The first form appears as \textbf{simple omissions}, where the model reconstructs most of the control flow correctly but omits the required boilerplate needed for compilation, such as headers or small structural definitions. In our experiments, this type of error became more frequent as the obfuscation strength increased; some models continued to produce code but omitted important elements, even though the control flow structure was preserved. The second form is \textbf{structural hallucination}, where the model outputs syntactically complete code but key data structures are malformed or only partially reconstructed, which can break type correctness in many functions. The third and most harmful form is \textbf{semantic hallucination}, in which the model incorrectly removes or edits control flow branches, altering program behavior, even if the resulting source code looks clean and well-formatted. Because these errors directly affect correctness and trust in the deobfuscation process, hallucination should be treated as a central factor in evaluating the reliability of LLM-based deobfuscation.

A concrete example of \textbf{simple omissions} hallucination comes from a sample program deobfuscated by DeepSeek-V2, as shown in figure \ref{figure5}. In this case, the right side of the figure presents the deobfuscated code produced by the model, while the left side shows the original program. The model correctly reconstructs the overall control flow and produces code that is close to the original, but it forgets to include the headers \texttt{\#include <stdio.h>} and \texttt{\#include <stdlib.h>}, and the \texttt{Node} structure is not defined correctly. As a result, the generated code does not compile in its initial form. However, once these hallucinated errors are corrected by adding the missing headers and fixing the structure definition, the program compiles and runs as expected, and the runtime outputs match the original behavior. This example illustrates how hallucination can appear even when the main control flow is preserved and shows that increasing obfuscation strength tends to push models toward more frequent omissions, structural errors, and, in more severe cases, semantic changes that are much harder to detect and repair.

Figure~\ref{figure6} presents a clearer example of \textbf{semantic hallucination} during LLM-based deobfuscation. In this case, the model produces code that is syntactically valid, readable, and easy to accept as correct, but it changes the original program behavior. The source function computes the height of a binary tree by taking the larger value between the left and right subtree heights. However, the hallucinated output replaces this logic with an additive form that sums the outputs of both recursive calls. As a result, the generated function no longer computes the tree height and instead behaves like a node-counting routine for the given example tree. This type of error is more serious than missing headers or incomplete type declarations because it is harder to detect by visual inspection and directly harms semantic correctness.

Although full code deobfuscation remains a very hard problem, our results show that the models in this study achieved high accuracy on the given dataset and behaved robustly and adaptively when deobfuscation was not needed. To test this, we ran an additional experiment in which we passed non-obfuscated code samples through the same CoT reasoning pipeline that we use for obfuscated inputs. The goal was to determine whether the models could detect that the code was already in a clean form and, therefore, avoid unnecessary changes. In the cases we examined, the models correctly recognized the absence of obfuscation and preserved the original code, rather than rewriting it or adding invented content. They did not alter the control flow structure, rename variables, or introduce spurious edits.

\section{Future Work}
After analyzing the use of CoT prompting for code deobfuscation, two main research questions remain: first, how can this approach be extended beyond opaque predicates and control-flow flattening to more advanced control-flow obfuscation techniques? and second, how can the methodology be improved to increase the accuracy of the deobfuscation process? We plan to build on CoT prompting, as it improves code explainability during the deobfuscation process by making the model show its reasoning step by step in a form that humans can read and review. This helps analysts review intermediate decisions, test the evidence supporting them, and identify errors more easily. We also plan to extend this line of work by studying self-consistency and Tree-of-Thought (ToT) prompting. Self-consistency may be useful for polymorphic and metamorphic malware because it can generate multiple reasoning paths and identify conclusions that remain stable even as the malware changes its visible form. ToT may also be valuable for malware that uses opaque predicates, as it can explore multiple execution paths and compare them in a structured way. This may help separate false or misleading branches from the real execution path. In addition, we plan to expand our study to include more obfuscation methods, particularly a broader range of control-flow and data obfuscation techniques. Finally, we aim to evaluate these methods on real malware samples, such as \textbf{Emotet}, to better understand how well the approach works in practical settings and to gain deeper insight into real malware behavior.

\section{Limitations}
This study has some limitations that should be considered when interpreting its findings. First, the work covers only a limited part of the code deobfuscation problem, examining opaque predicates, control flow flattening, and their combination, using only two open-source obfuscation tools: Tigress and O-LLVM. Open source tools were used for reproducibility, but there is a lack of comprehensive Open source obfuscation tools for research. Second, the method's reliability remains an important concern. The results show that CoT based deobfuscation can still produce hallucinations, meaning outputs that appear correct but contain errors such as missing headers, incomplete data structures, or larger semantic changes that alter program behavior. The method also becomes less stable when the number of opaque branches increases, when layered obfuscation is applied, or when the original control flow graph is more complex. In addition, the study shows that explicit guidance on reasoning in the prompt improves performance. This suggests that part of the reported improvement may come from prompt design and task specific examples, rather than from CoT alone. For this reason, the findings should be viewed within these limits, and broader validation is still needed before strong general claims can be made about the effectiveness of this approach.

\section{Conclusion}
Code deobfuscation remains a difficult and open problem in cybersecurity. However, our results show that large language models can make meaningful progress when appropriately guided. Using a structured CoT prompting strategy and leveraging the models’ basic code-explainability skills, we find that they can deobfuscate both low-level code, such as LLVM-IR, and high-level C programs. In this setting, the models can walk through the code step by step, track identifiers and variables, and detect common obfuscation constructs, including opaque predicates and unreachable branches. At the same time, all models still show some degree of hallucination, and their performance degrades as we increase the number of opaque predicate branches or apply multiple layers of obfuscation. A key piece of evidence comes from experiments where we explicitly embed logical reasoning steps inside the CoT prompt; for example, short explanations of why a given predicate is always true or always false, as illustrated in figure \ref{figure1}. In these cases, the models follow a more coherent reasoning path, remove more opaque branches correctly, and produce deobfuscated code with higher structural and semantic scores, confirming that explicit reasoning guidance is an effective way to strengthen LLM-based deobfuscation.

\section*{Acknowledgment}

The authors would like to express their sincere appreciation to the students of the KAI2 Lab for their invaluable support, collaboration, and engagement throughout the course of this research. Their contributions significantly enhanced the quality and progress of this work. The authors also gratefully acknowledge the UMBC Cybersecurity Institute for its continued support and resources, which were instrumental in facilitating this research.

\bibliographystyle{IEEEtran}
\bibliography{biblio}

\end{document}